\documentclass[acmsmall,screen]{acmart}

\usepackage{xspace}
\usepackage{pifont}
\usepackage{wrapfig}
\usepackage{tabularx}
\usepackage{footnote}
\usepackage{makecell}
\usepackage{multirow}
\usepackage{booktabs}
\usepackage{graphicx}
\usepackage{textcomp}
\usepackage{enumerate}
\usepackage{todonotes}
\usepackage{subcaption}
\usepackage{algorithmic}
\usepackage{color, xcolor}
\usepackage{amsmath,amsthm}
\usepackage[most]{tcolorbox}

\newcommand{\ie}{\textit{i.e.,}\xspace}
\newcommand{\eg}{\textit{e.g.,}\xspace}
\newcommand{\etc}{\textit{etc.}\xspace}
\newcommand{\etal}{\textit{et al.}\xspace}

\newcommand{\tabref}[1]{Table~\ref{#1}\xspace}
\newcommand{\figref}[1]{Fig.~\ref{#1}\xspace}
\newcommand{\secref}[1]{Section~\ref{#1}\xspace}

\newcommand{\toolname}{{\sc PersonaTester}\xspace}
\newcommand{\ps}[1]{{\sc Persona {#1}}\xspace}

\begin{document}

\sloppy

\title{Towards Automated Crowdsourced Testing via Personified-LLM}

\author{Shengcheng Yu}
\affiliation{\department{School of Computation, Information and Technology, Institute for Advanced Study, Heilbronn Data Science Center, Munich Data Science Institute}\institution{Technical University of Munich}\city{Heilbronn}\country{Germany}}
\email{shengcheng.yu@tum.de}
\orcid{0000-0003-4640-8637}

\author{Yuchen Ling}
\affiliation{\institution{State Key Laboratory of Novel Software Technology, Nanjing University}\city{Nanjing}\country{China}}
\email{yuchenling@smail.nju.edu.cn}
\orcid{0000-0000-0000-0000}

\author{Chunrong Fang}
\affiliation{\institution{State Key Laboratory of Novel Software Technology, Nanjing University}\city{Nanjing}\country{China}}
\email{fangchunrong@nju.edu.cn}
\orcid{0000-0002-9930-7111}

\author{Zhenyu Chen}
\affiliation{\institution{State Key Laboratory of Novel Software Technology, Nanjing University}\city{Nanjing}\country{China}}
\email{zychen@nju.edu.cn}
\orcid{0000-0002-9592-7022}

\author{Chunyang Chen}
\affiliation{\department{School of Computation, Information and Technology, Munich Data Science Institute, Heilbronn Data Science Center, Fortiss}\institution{Technical University of Munich}\city{Heilbronn}\country{Germany}}
\email{chun-yang.chen@tum.de}
\orcid{0000-0003-2011-9618}

\ccsdesc[500]{Software and its engineering~Software testing and debugging}

\begin{abstract}
The rapid proliferation and increasing complexity of software demand robust quality assurance, with graphical user interface (GUI) testing playing a pivotal role. Crowdsourced testing has proven effective in this context by leveraging the diversity of human testers to achieve rich, scenario-based coverage across varied devices, user behaviors, and usage environments. In parallel, automated testing, particularly with the advent of large language models (LLMs), offers significant advantages in controllability, reproducibility, and efficiency, enabling scalable and systematic exploration. However, automated approaches often lack the behavioral diversity characteristic of human testers, limiting their capability to fully simulate real-world testing dynamics. To address this gap, we present \toolname, a novel personified-LLM-based framework designed to automate crowdsourced GUI testing. By injecting representative personas, defined along three orthogonal dimensions: testing mindset, exploration strategy, and interaction habit, into LLM-based agents, \toolname enables the simulation of diverse human-like testing behaviors in a controllable and repeatable manner. Experimental results demonstrate that \toolname faithfully reproduces the behavioral patterns of real crowdworkers, exhibiting strong intra-persona consistency and clear inter-persona variability (117.86\% -- 126.23\% improvement over the baseline). Moreover, persona-guided testing agents consistently generate more effective test events and trigger more crashes (100+) and functional bugs (11) than the baseline without persona, thus substantially advancing the realism and effectiveness of automated crowdsourced GUI testing.
\end{abstract}

\keywords{Software Testing, Crowdsourced Testing, LLM, LLM Personification}

\setcopyright{cc}
\setcctype{by-nc-nd}
\acmJournal{PACMSE}
\acmYear{2026}
\acmVolume{3}
\acmNumber{FSE}
\acmArticle{FSE166}
\acmMonth{7}
\acmDOI{10.1145/3808173}
\acmSubmissionID{fse26mainb-p1153-p}
\received{2025-09-02}
\received[accepted]{2026-03-24}

\maketitle

\section{Introduction}

With the increasing complexity and interactivity of modern software systems, ensuring robust quality assurance has become more critical than ever \cite{yu2023vision}. Among various testing paradigms, crowdsourced testing has emerged as a powerful manual testing strategy that leverages the collective intelligence of a distributed pool of human testers (\ie crowdworkers) recruited via online platforms to evaluate software quality \cite{gao2019successes, wang2022context}. As a scalable extension of traditional manual testing, crowdsourced testing enables rapid recruitment of participants from diverse backgrounds, devices, and regions, thereby capturing a wide range of real-world user behaviors. This diversity brings significant benefits in terms of scenario coverage, functional exploration, and defect discovery, especially for systems with complex or user-centric interfaces \cite{yu2019crowdsourced}. Crowdworkers contribute heterogeneous interaction styles shaped by differences in culture, education, domain knowledge, testing mindset, and input preferences, helping trigger bugs that are difficult to identify through conventional testing. Widely adopted in both industry and academia, platforms such as uTest, Baidu CrowdTest, and MoocTest have facilitated thousands of test campaigns across web, mobile, and desktop software \cite{wang2020context, wang2022intelligent}.

A key strength of crowdsourced testing lies in its ability to amplify behavioral diversity. Crowdworkers contribute heterogeneous interaction styles shaped by differences in education, culture, domain knowledge, testing mindset, and input preferences, resulting in rich and varied testing traces. This diversity improves functional and scenario-level coverage and allows for the detection of edge-case bugs and usability issues often missed by conventional in-house testing. For example, alpha and beta testing in game development frequently rely on crowds to uncover region- or device-specific problems. Prior studies have shown that crowdsourced test reports frequently capture real-world failure scenarios that are difficult to replicate through automated testing alone \cite{fang2024enhanced, huang2020quest}.

However, the manual nature of crowdsourced testing also introduces inherent limitations \cite{yu2024semi, fang2024enhanced, yu2024effective}. Coordination overhead, result variability, lack of reproducibility, and high labor costs make it difficult to scale consistently, particularly across iterative development cycles. While diversity is its greatest strength, it also poses challenges in maintaining test reliability, coverage traceability, and feedback consolidation. These issues have led to increasing interest in evolving crowdsourced testing into a more scalable, automated paradigm, one that can simulate the breadth of human behavior while minimizing the downsides of human involvement. Recent advancements in automated testing \cite{su2017guided, pan2020reinforcement, li2017droidbot, wang2020combodroid, amalfitano2012using, li2019humanoid}, particularly those driven by large language models (LLMs) \cite{liu2024make}, offer opportunities in this regard. LLMs possess strong reasoning capabilities and can generate semantically meaningful test actions with minimal manual supervision. However, conventional LLM-based agents often lack the behavioral variability and interaction richness exhibited by real human testers \cite{chen2025standing, liu2024unblind, liu2024make}, making them less effective at replicating the diverse exploration paths in crowdsourced testing. Most automated approaches adopt fixed strategies, resulting in repetitive and narrow test behaviors that fail to reflect real-world usage diversity.

To address this gap, we propose \toolname, a novel framework that evolves traditional crowdsourced testing into an automated crowdsourced testing paradigm by integrating crowdworker personas into LLM-based testing workflows. \toolname harnesses human behavioral diversity by explicitly modeling three orthogonal dimensions: testing mindset, exploration strategy, and interaction habit, derived from large-scale analysis of real crowdsourced test traces. These personas serve as lightweight cognitive profiles that guide LLM-driven decision-making, allowing each automated test instance to reflect unique and realistic testing behavior. By combining human-inspired diversity with LLM-enabled scalability and control, \toolname preserves the strengths of both human intelligence and artificial intelligence. Instead of simply integrating human domain knowledge through mechanisms such as knowledge graphs (\ie app-side knowledge), the human-side information, such as personality traits or interaction preferences of testers, should also be considered for better diversity. In essence, \toolname reimagines crowdsourced testing not as a purely manual process \cite{yu2024practical}, but as a synergistic collaboration between the collective behavioral patterns of human testers and the reasoning power of LLMs. This paradigm shift unlocks a new direction for scalable, reproducible, and human-like software testing, bringing the benefits of crowd diversity into automated test generation without relying on human labor.

\toolname introduces a structured yet flexible framework for personified LLM-based GUI testing, consisting of four tightly integrated components: LLM personification modeling, GUI state understanding, LLM-based decision-making, and operation execution with feedback validation. At the core of the framework lies its personification mechanism, which models the LLM testing agent's behavior using empirically grounded personas defined along three orthogonal dimensions: Testing Mindset, Exploration Strategy, and Interaction Habit. These dimensions capture cognitive orientation (\eg sequential vs. divergent logic), interaction preferences (\eg clicking, input-driven actions, or core function prioritization), and input styles (\eg short valid, long boundary, or invalid values). Through empirical analysis of real-world crowdsourced testing reports, we construct nine representative personas that reflect diverse testing behaviors observed in practice. Personas are explicitly injected into the LLM's prompt, ensuring that decisions of LLM agents remain coherent, reproducible, and aligned with realistic behavioral patterns.

During each testing session, the LLM agent operates iteratively. The process begins with robust GUI perception, where screenshots are processed through a hybrid pipeline combining computer vision techniques and multimodal large language models (MLLMs) to extract textual, structural, and spatial widget information. A filtering mechanism removes static or non-interactable elements and identifies transient components (\eg drop-down menus), after which the GUI is textualized into a JSON representation that supports precise downstream reasoning. LLM agents then receive contextual prompts containing the structured GUI state, test history, and assigned persona, and first generate a high-level testing intent (\eg ``modify alarm setting'') to clarify the reasoning objective. This is followed by the generation of a specific operation aligned with both the intent and the agent's interaction habit. This two-step process promotes interpretability and consistency, allowing for semantically rich and behaviorally meaningful test sequences. The generated operations are executed based on the mapped GUI coordinates. Following execution, \toolname validates the effect of the operation through intent-based state checking and visual-semantic bug detection using MLLMs. This closed-loop design, linking reasoning, execution, and validation, enables intelligent, autonomous GUI testing that authentically simulates the diverse strategies of human crowdworkers.

Experimental results demonstrate that the proposed \toolname effectively imitates human crowdworkers, thus automating and enhancing the crowdsourced testing. \toolname can achieve 117.86\% -- 126.23\% improvement in intra-persona consistency and distinct inter-persona variability over the baseline. Specifically, agents with the same persona consistently exhibit highly similar exploration trends, closely aligning with their assigned persona characteristics, while agents with different personas demonstrate notably varied exploration behaviors. Furthermore, by injecting personas representing diverse testing mindsets, exploration strategies, and interaction habits, persona-guided agents generate testing events more effectively, closely correlated with persona attributes, and consistently surpass the performance of agents without persona injection. Additionally, persona-guided agents identify more crash bugs (100+) and functional bugs (11) compared to the baseline, highlighting the effectiveness and practicality of \toolname in automating and enhancing crowdsourced GUI testing.

The noteworthy contributions of this paper can be concluded as follows:

\begin{itemize}
\item This paper presents \toolname, a novel framework that practically enables personified LLM agents to simulate diverse and realistic human-like GUI testing behaviors, which is the first work to apply the \textbf{``persona''} concept to software testing.
\item This paper introduces a structured three-dimensional persona schema that systematically models testing mindsets, exploration strategies, and interaction habits for test exploration.
\item Empirical experiments show persona-guided LLM agents enhance test diversity and effectiveness, outperforming the non-personified baseline in bug triggering.
\end{itemize}

\section{Preliminary Study}

To ground the design of \toolname in real-world testing practices, we first conduct an investigation into a large corpus of crowdsourced GUI testing reports. This analysis allows us to abstract common behavioral patterns of human testers into structured persona dimensions and configurations, which serve as the foundation of our personified testing framework. We then present an illustrative example that demonstrates how different persona-guided agents explore the same testing task in distinct ways. These motivate the need for simulating diverse human-like testing behaviors, highlighting the potential of integrating persona modeling into automated crowdsourced testing.

\subsection{Persona Investigation from Real Crowdsourced Testing}

To ensure the realism and representativeness of our persona modeling, we conduct an empirical investigation into real-world crowdsourced GUI testing behaviors. We randomly sample 1,500 GUI exploration traces from a public dataset of crowdsourced test reports drawn from one of the most widely used and academically studied platforms\footnote{Anonymized for the double-anonymous review criteria.} \cite{yu2021prioritize, yu2023mobile, fang2024enhanced}. This dataset includes approximately 23,000 reports spanning over 50 software systems and involving more than 1,100 distinct crowdworkers. Notably, the testing process on this platform does not assign crowdworkers fixed tasks; instead, they are only provided with requirement documentation describing the app's functionality. This open-ended testing setup encourages testers to explore autonomously and naturally, allowing individual behavioral traits to be more clearly expressed.

Through qualitative analysis of the sampled traces, we identify three recurring and discriminative aspects of crowdworker behavior: their testing mindset (\eg whether their exploration is sequential and methodical or divergent and opportunistic), exploration strategy (\eg emphasis on clickable elements, input interactions, or core functional paths), and interaction habit (\eg the nature and length of input text during test operations). These three dimensions describe the user interactions of crowdworkers from the high-level mindset, middle-level strategy, and concrete low-level habits, respectively. Such dimensions not only recur consistently across traces but also align with prior work on human testing behavior and software testing diversity. Together, they offer a compact yet expressive space for modeling crowdworker variability in test generation.

We then manually annotate each of the 1,500 traces across the three dimensions. The annotation is performed by three authors independently and finalized through consensus. Among the 18 possible combinations (2 $\times$ 3 $\times$ 3), we identify 9 persona configurations with frequencies exceeding 1\%, ranging from 2.27\% to 21.80\%. These 9 configurations together account for 95.40\% of all observed traces, while the remaining combinations fall below the 1\% threshold. We therefore select these empirically grounded 9 personas in \toolname, balancing behavioral diversity with real-world representativeness. Notably, this subset satisfies pairwise coverage across all dimensions \cite{nie2011survey}, ensuring that each attribute is represented in multiple combinations. This design supports comprehensive evaluation of how different persona traits influence test behavior and defect discovery, while preserving theoretical completeness and practical feasibility.

\subsection{Illustrative Example}
\label{sec:mv}

\begin{figure}[!h]
\centering
\includegraphics[width=0.9\linewidth]{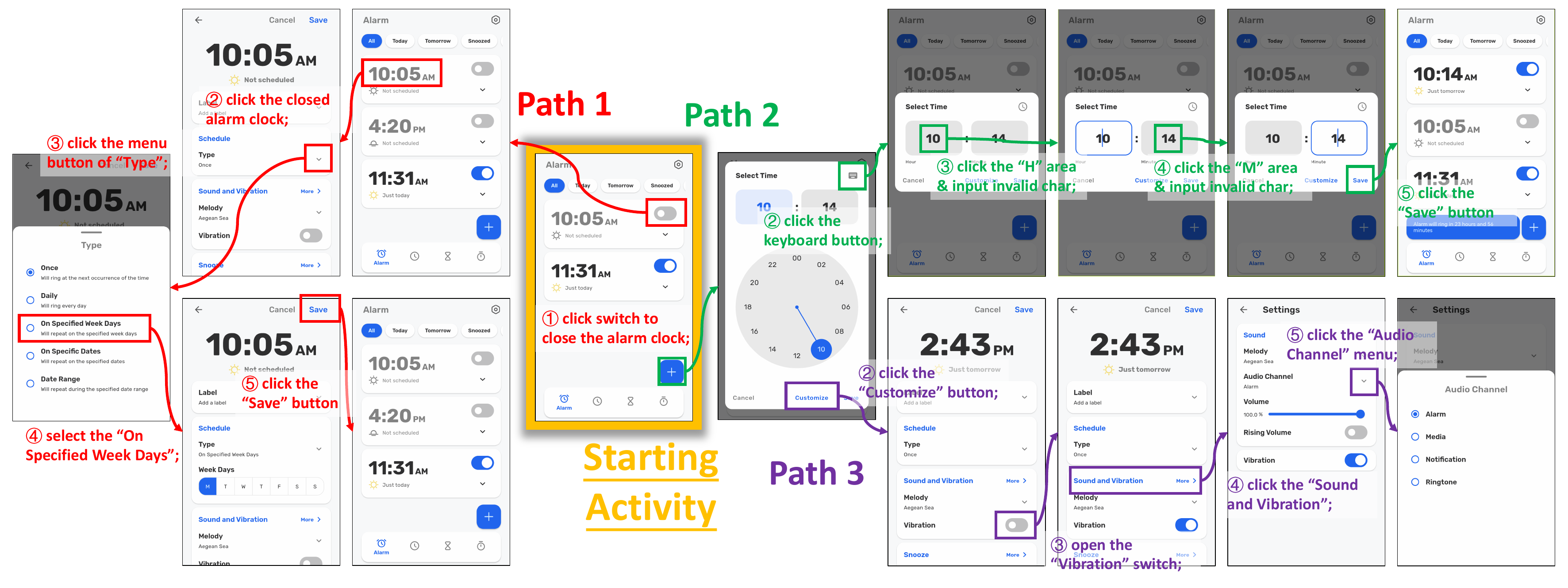}
\vspace{-10pt}
\caption{Motivating Example: Exploration Trace of Personified-LLM Agents with Different Personas}
\vspace{-10pt}
\label{fig:mv}
\end{figure}

To demonstrate the behavioral diversity and practical testing effectiveness enabled by persona-guided LLM-based GUI testing, we present an illustrative example involving three testing sessions conducted by \toolname configured with distinct personas: \ps{B}, \ps{C}, and \ps{E} (design details of persona-guided LLM agents in \secref{sec:persona}). The specific task given to the LLM agents is to explore the alarm clock management function of the given app. Each persona defines a unique combination of testing mindset, exploration strategy, and interaction habit, which in turn drives the agent's behavior during test execution (see \figref{fig:mv}). This example highlights how the personas lead to differentiated exploration paths that correspond to the variability of human testers in crowdsourced testing and reveal potential issues from diverse perspectives.

\begin{center}\scalebox{0.9}{\framebox{\parbox{0.9\linewidth}{
\textbf{\ps{B} (Path 1)}: Testing Mindset A. sequential \& coherent, Exploration Strategy b. core function focused, Interaction Habit ii. valid \& short input \\
\textbf{\ps{C} (Path 2)}: Testing Mindset A. sequential \& coherent, Exploration Strategy c. input oriented, Interaction Habit iii. invalid input \\
\textbf{\ps{E} (Path 3)}: Testing Mindset B. divergent \& non-linear, Exploration Strategy a. click oriented, Interaction Habit ii. valid \& short input
}}}\end{center}

In the first case (Path 1 in \figref{fig:mv}), the agent equipped with \ps{B} exhibits a sequential and coherent testing mindset, a core-function-focused exploration strategy, and a habit of providing short and valid inputs. This agent follows a task-directed and systematic exploration pattern, beginning by clicking the switch to disable an existing alarm clock. It then accesses the editing page of the deactivated alarm, navigates to the menu for configuring the alarm type, selects the ``On Specified Week Days'' option, and finally clicks the ``Save'' button to confirm the changes. This exploration not only aligns well with the defined behavioral profile of \ps{B} but also leads to the discovery of a \textbf{functional bug}: after editing a closed alarm and modifying its type, the updated alarm cannot be reopened or deleted. This bug is triggered specifically by the sequence of opening a closed alarm, changing its type, and saving the changes, a condition that is unlikely to be encountered through standard automated testing policies that lack contextual and purpose-driven reasoning. The bug discovery here illustrates the importance of incorporating persona-driven exploration logic into the testing process.

In contrast, the agent with \ps{C} (Path 2 in \figref{fig:mv}) also follows a sequential and coherent testing mindset but is configured with an input-oriented exploration strategy and a habit of generating short, invalid inputs. The agent initiates its exploration by clicking the ``+'' button to add a new alarm, followed by interactions with the time-setting interface. It attempts to enter invalid characters in both the hour and minute input areas and proceeds to click the ``Save'' button. This trace represents a valuable testing trajectory that assesses the robustness of the input-handling logic. Such behavior reflects the persona's orientation toward identifying improper input scenarios, a type of testing often missed by traditional automated tools that lack a notion of adversarial interaction patterns.

The third trace (Path 3 in \figref{fig:mv}), driven by \ps{E}, reflects a markedly different behavioral style. With a divergent and non-linear testing mindset, a click-oriented exploration strategy, and a preference for short and valid interactions, the agent initiates testing by clicking the ``+'' button but quickly deviates into secondary settings. It selects the ``Customize'' option, enables the ``Vibration'' toggle, accesses the ``Sound and Vibration'' settings page, and opens the ``Audio Channel'' configuration menu. This behavior typifies a curious, exploratory user and enables the testing of GUI states beyond primary functionalities. This exploration contributes much to coverage diversity by exercising peripheral pathways that structured exploration strategies might overlook.

Together, these three exploration traces illustrate how different personas guide LLM-based agents toward distinct and meaningful testing behaviors. \ps{B} uncovers a critical functional bug through systematic task-focused navigation, \ps{C} tests boundary conditions of input validation, and \ps{E} exercises lesser-explored GUI pathways through non-linear interaction. Crucially, only with persona-guided LLM exploration can such behavioral differentiation be achieved. Traditional automated GUI testing frameworks, relying on uniform and deterministic strategies, lack the flexibility to simulate diverse user perspectives. In contrast, the persona-driven design of \toolname allows LLM-based testing agents to adopt human-like reasoning patterns and behavioral intents, enabling realistic, adaptive, and effective GUI exploration. This set of examples underscores the practical value of integrating personified LLM agents into crowdsourced GUI testing, enhancing the automation of crowdsourced GUI testing while preserving the richness and variability of human tester behavior.

\section{Methodology}

\begin{figure}[!h]
\centering
\vspace{-0.3cm}
\includegraphics[width=0.92\linewidth]{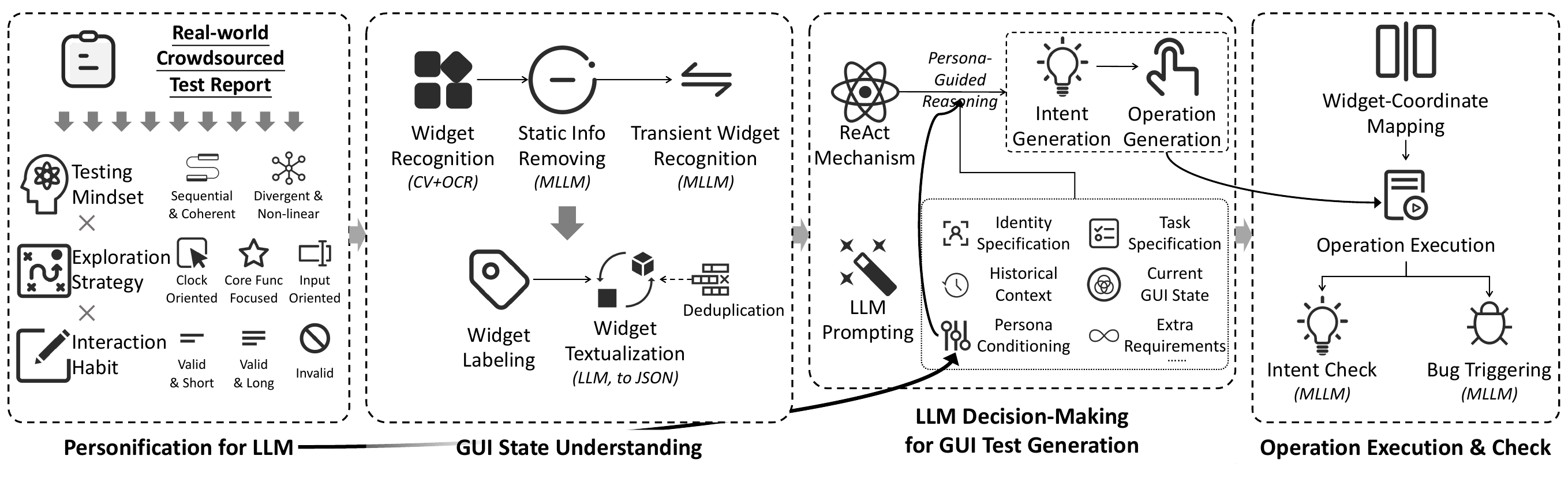}
\vspace{-0.3cm}
\caption{Overview of \toolname Workflow}
\vspace{-0.5cm}
\label{fig:framework}
\end{figure}

\toolname integrates human-like behavioral diversity to automate crowdsourced testing through a structured pipeline driven by personified LLM agents (\figref{fig:framework}). We first formalize personas, representing the testing mindset, exploration strategy, and interaction habit, respectively. Based on empirical analysis of real-world crowdsourced testing behaviors, we select nine representative personas to guide LLM behavior. The iterative process then begins with GUI state understanding, where app GUI screenshots are processed using CV and multimodal models to extract and persist structured widget representations. Test generation proceeds via a two-step prompting process, intent formulation followed by persona-aligned operation generation, enhancing interpretability and consistency. Finally, the operation is executed, and its outcome is verified through intent checking and MLLM-based bug triggering. This end-to-end framework combines the exploration automation with the behavioral realism of crowdworkers, offering a novel paradigm for GUI testing. 

\subsection{Personification for LLM}
\label{sec:persona}

To simulate the behavioral diversity observed in real-world crowdsourced testing, we adopt a persona-based modeling approach, drawing inspiration from established concepts in human-computer interaction (HCI) and conversational AI. In general, a persona refers to a fictitious yet data-driven representation of a user archetype, commonly used to model user needs, behaviors, and goals in system design and evaluation \cite{grudin2002personas}. In the context of LLM applications, personas have been used to shape response style, reasoning logic, and domain alignment by embedding role-specific attributes into prompts \cite{hu2024quantifying}. Building on these foundations, we define a persona in our framework as an abstract tester profile that encapsulates specific behavioral tendencies and decision-making preferences frequently observed in crowdworkers. Guided by empirical analysis of real crowdsourced test reports, we introduce a structured and interpretable schema that decomposes human testing behavior into three orthogonal dimensions: Testing Mindset (cognitive orientation during exploration), Exploration Strategy (preferred interaction targets), and Interaction Habit (input generation style). This structured formulation supports the systematic instantiation of diverse, personified LLM-based testing agents, each emulating a distinct and realistic tester profile to enrich automated GUI testing with human-like behavioral diversity. This is the first work to apply \textbf{``persona''} to software testing.

\begin{figure}[!h]
\centering
\vspace{-10pt}
\includegraphics[width=\linewidth]{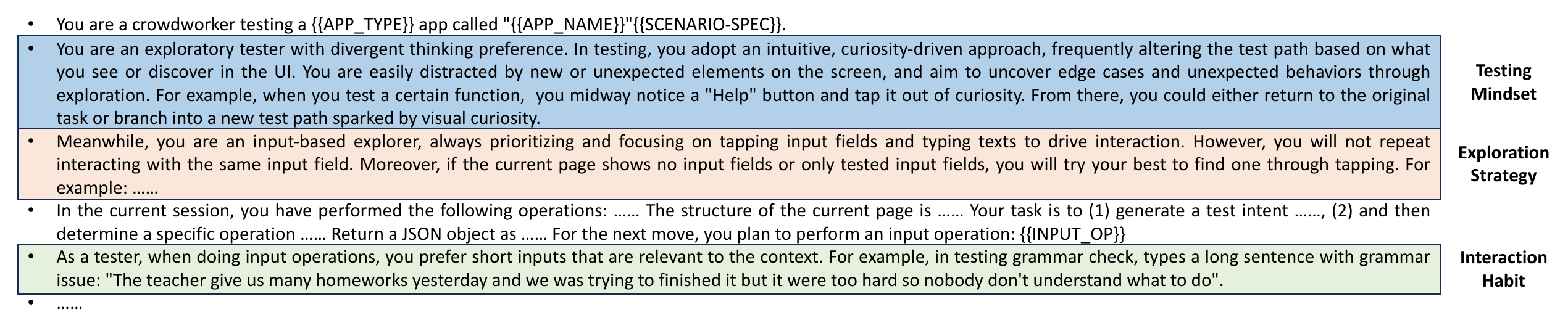}
\vspace{-0.8cm}
\caption{Example of Prompt for LLM Personification (\ps{F})}
\vspace{-10pt}
\label{fig:prompt}
\end{figure}

Testing Mindset defines the high-level cognitive orientation of the agent, which influences the structure and logic of the exploration process. We model this using two attributes: \textit{A. sequential\_and\_coherent}, representing crowdworkers who follow a structured and linear exploration path with a goal-directed flow; and \textit{B. divergent\_and\_non-linear}, denoting crowdworkers who adopt more scattered, curiosity-driven paths. These attributes abstract the principal thinking patterns observed in crowdsourced testing sessions and capture a wide range of human cognitive styles.

Exploration Strategy specifies the tactical preference of the agent when choosing where to interact in the GUI. It consists of three commonly observed testing tendencies: \textit{a. click\_oriented}, favoring general interaction with clickable widgets; \textit{b. core\_function\_focused}, prioritizing features central to app functionality; and \textit{c. input\_oriented}, concentrating on widgets that accept user input. These categories are derived from an empirical investigation of the previous study \cite{yu2023mobile} that most user interactions in crowdsourced GUI testing (>90\%) fall into click and input actions, and that functional goal orientation is a primary heuristic for manual testing.

Interaction Habit captures the style of input generation during testing. Since input operations are highly sensitive to the content and type of values, we model input behavior with three distinctive attributes:  \textit{i. valid\_and\_short}, corresponding to standard user entries; \textit{ii. valid\_and\_long}, representing stress testing via lengthy or extreme values; and \textit{iii. invalid}, focusing on edge-case or error-triggering input. This dimension reflects realistic variance in user interaction behavior that can significantly influence the app execution path and potential for defect discovery.

\begin{table}[!h]
\vspace{-10pt}
\caption{Configuration of Persona-Guided LLM Agent}
\vspace{-10pt}
\label{tab:agent}
\scalebox{0.7}{\begin{tabularx}{\textwidth}{c|*{2}{>{\centering\arraybackslash}X}|*{3}{>{\centering\arraybackslash}X}|*{3}{>{\centering\arraybackslash}X}}

\toprule
\multirow{2}{*}{\makecell{Persona-Guided \\ Agent}} & 
\multicolumn{2}{c|}{Testing Mindset} & 
\multicolumn{3}{c|}{Exploration Strategy} & 
\multicolumn{3}{c}{Interaction Habit} \\ \cmidrule{2-9}

& A & B & a & b & c & i & ii & iii \\ \midrule

\ps{A} & \checkmark &  & \checkmark &  &  & \checkmark &  &  \\
\ps{B} & \checkmark &  &  & \checkmark &  &  & \checkmark &  \\
\ps{C} & \checkmark &  &  &  & \checkmark &  &  & \checkmark \\
\ps{D} &  & \checkmark &  & \checkmark &  &  &  & \checkmark \\
\ps{E} &  & \checkmark & \checkmark &  &  &  & \checkmark &  \\
\ps{F} &  & \checkmark &  &  & \checkmark & \checkmark &  &  \\
\ps{G} & \checkmark &  & \checkmark &  &  &  &  & \checkmark \\
\ps{H} &  & \checkmark &  & \checkmark &  & \checkmark &  &  \\
\ps{I} & \checkmark &  &  &  & \checkmark &  & \checkmark &  \\\bottomrule

\end{tabularx}

}\end{table}

To instantiate personified LLM agents, we define a persona as a tuple: $\text{Persona} = \langle m, s, h \rangle,\quad m \in \{A, B\},\ s \in \{a, b, c\},\ h \in \{i, ii, iii\}$. Each tuple encodes a complete behavioral profile used to prompt and constrain the LLM agent's test generation behavior. This formalization enables the instantiation of personified agents with well-defined behavioral priors, facilitating controlled experimentation and reliable behavior emulation in automated GUI testing. By conditioning the LLM agents with these persona tuples during prompt construction, \toolname achieves consistent, reproducible, and realistic simulation of diverse human testing behaviors, significantly advancing the integration of human factors into automated GUI testing.

\subsection{GUI State Understanding}

A critical foundation of \toolname lies in its capability to interpret the app GUI state. This interpretation must not only capture the visible structure and semantics of each state, but also dynamically reflect transient states and actionable GUI elements essential for persona-guided reasoning. To this end, we design a multi-stage GUI understanding method that combines traditional Computer Vision (CV) techniques with multimodal large language models (MLLMs). This hybrid strategy ensures both precision and generalizability across varied GUI contexts.

The GUI state understanding process begins with widget recognition, which uses a combination of traditional CV and OCR techniques to extract raw GUI elements from the GUI states \cite{yu2024effective}. This step identifies widget boundaries and textual content while preserving layout information. Importantly, we do not directly rely on end-to-end MLLMs for this step, as such models cannot perform well on complex GUI layouts without extensive fine-tuning and often miss structural information critical for downstream reasoning \cite{su2025automated}. Following initial recognition, we perform static infomation removal using an MLLM-based filter to eliminate GUI elements that do not contribute to meaningful interaction \cite{liu2024unblind}. These include persistent status bars, decorative elements, and non-functional text. This step improves the signal-to-noise ratio of the GUI state and prevents the agent from being distracted by irrelevant components. Next, transient widget recognition is applied to identify context-sensitive GUI elements, such as temporary modals, dropdown menus, or pop-up windows. These transient widgets are often overlooked by static parsing but play a significant role in determining the current GUI state. The MLLM is used again in this phase, leveraging its visual-linguistic reasoning to distinguish ephemeral elements from the persistent layout.

The refined set of GUI widgets then undergoes widget labeling, where each GUI element is categorized (\eg button, input field, toggle switch) and associated with a semantic descriptor. This facilitates downstream interpretation by the LLM agent and ensures the agent receives structured, labeled input aligned with its persona-guided decision-making. Then, in the widget textualization step, we convert the labeled GUI state into a structured JSON representation using an LLM. Each GUI widget is expressed as a key-value pair that captures its type, content, and potential interactions. This JSON representation serves as the semantic abstraction of the GUI state, enabling consistent interpretation across different LLM prompts. The use of LLMs in this stage allows for robust natural language grounding, especially when widget labels or contexts are ambiguous. Finally, to optimize performance and avoid redundant processing, we incorporate widget persistence. Screenshots with high visual similarity (cosine similarity above 0.99, which is determined according to existing studies \cite{yu2021prioritize, yu2023vision}) are detected and cached, allowing the system to reuse previously parsed JSON representations. This not only reduces computational overhead but also ensures consistent state interpretation in recurring GUI contexts.

\toolname balances the strengths of CV algorithms with the flexible semantic reasoning capabilities of (M)LLMs. By explicitly addressing both persistent and transient elements, and transforming visual information into a structured, persona-comprehensible format, our GUI state understanding module establishes a robust foundation for intelligent and diverse test exploration.

\subsection{LLM Decision-Making for GUI Test Generation}

As an automated crowdsourced testing approach, \toolname features a structured decision-making pipeline, designed to generate semantically meaningful and persona-guided GUI testing behaviors. This process unfolds iteratively, drawing inspiration from the ReAct paradigm, where each step dynamically determines the next test action based on the current context, including GUI state, testing history, and persona configuration. The decision-making prompt is constructed in a modular fashion. It begins with explicit identity and task specifications to inform the LLM about the target app and the scenario-specific testing goal. Crucially, the prompt embeds the persona profile, particularly the testing mindset and exploration strategy, which steers the reasoning process. For example, a persona with a sequential and task-focused profile is guided to prioritize central GUI elements in a top-down logical order, whereas a divergent persona may explore peripheral or less conventional interface paths. The prompt also integrates the historical interaction context and the structured GUI representation generated during the perception phase, promoting coherence and reducing semantic redundancy across operations.

Built upon this contextual foundation, \toolname adopts a two-stage reasoning and action process. First, the LLM performs test intent generation, articulating a high-level goal such as ``attempt to toggle notification settings''. This semantic layer enhances interpretability, aligns with persona logic, and guides hidden element discovery where necessary. Next, intent-based test generation is carried out, producing a concrete GUI operation aligned with both the intended goal and the persona's interaction habit. For instance, personas inclined toward boundary testing may issue long or invalid inputs, while others may favor short and valid entries. When no immediate interaction target is detected, the system proactively searches for hidden or transient widgets (\eg drop-down menus), leveraging the generated intent as a guide. This decoupled design, reasoning followed by action, yields multiple advantages over single-step approaches. It improves behavioral modularity, enhances debugging and traceability, and ensures greater consistency across testing sessions. Additionally, it supports tasks such as intent-alignment checking and widget prioritization.

To ensure robustness and behavioral fidelity, the prompt design further incorporates safeguards against repeated operations, enforces persona-aligned behavior, and prioritizes transient widgets to prevent missed ephemeral elements. The LLM's output is structured, comprising the test intent, target widget reference, action type, optional parameters (\eg text or scroll direction), and a summary for downstream traceability. In summary, this decision-making framework enables \toolname to generate interpretable, diverse, and strategically guided GUI testing behaviors. By integrating persona modeling, context awareness, and a ReAct-style reasoning-action mechanism, the system effectively simulates real-world crowdworker behavior while maintaining automation advantages such as stability, reproducibility, and semantic clarity.

\subsection{Operation Execution \& Check}

Once a test operation is generated, \toolname proceeds to execute and validate it. This stage ensures not only the actual execution of the operation but also the semantic verification of its intent and the detection of potential GUI-related defects. The process begins by translating the abstract target widget and associated operation into precise screen coordinates, using the structured GUI state representation constructed during the perception phase. This coordinate-level mapping guarantees compatibility with low-level GUI drivers, enabling accurate and device-agnostic interaction. The specified operation, such as tapping, text input, or scrolling, is then carried out on the device, after which the resulting screen is captured and passed to the validation pipeline.

The validation consists of two complementary procedures. First, an intent check verifies whether the executed operation achieves its intended effect. This is performed using an MLLM, which compares the expected outcome, defined by the generated intent, with the updated GUI state. This semantic validation provides a lightweight mechanism for assessing whether the app responds as expected. Second, a bug detection mechanism analyzes the post-operation interface for anomalies, including visual layout inconsistencies, missing or broken widgets, and signs of failure such as error messages or unexpected UI resets. By utilizing visual-linguistic reasoning, the approach remains robust across diverse interface designs and application scenarios. This design enables \toolname to conduct intelligent GUI exploration, while continuously assessing the effectiveness of diverse testing behaviors.

\section{Experiment}

To evaluate \toolname, we conduct a comprehensive study covering key aspects of personified-LLM-based crowdsourced testing. Our experiments investigate whether LLM agents guided by personas can simulate human-like testing behaviors and enhance automated crowdsourced testing. We assess behavioral consistency and diversity across different personas (RQ1), the effectiveness of generated test events in performing meaningful interactions (RQ2), and the agents' ability to trigger both crash and functional bugs (RQ3). The evaluation spans a diverse set of real-world apps and tasks, using both quantitative metrics and user study analyses to validate the findings.

\subsection{Experimental Setting}

\subsubsection{Research Questions (RQ)}

To evaluate \toolname in automated crowdsourced testing, we design a set of research questions (RQs) targeting key aspects of the framework. These RQs focus on assessing the behavioral patterns of persona-guided LLM agents, their test generation effectiveness, and their ability to trigger bugs, core factors that reflect the capacity to simulate human-like testing behaviors and improve the diversity and quality of GUI testing.

\textbf{RQ1} examines the \textbf{exploration trends} of \toolname, referring to whether persona-guided testing leads to consistent or distinguishable behavioral patterns. We evaluate intra-persona consistency (RQ1.1) by checking if the same persona produces stable exploration traces across runs, and inter-persona variability (RQ1.2) by comparing behaviors across different personas on the same task. To further assess alignment, we conduct a user study (RQ1.3) where participants rate how well each behavior reflects the corresponding persona dimensions.

\textbf{RQ2} evaluates the \textbf{test generation effectiveness}, indicating whether persona-guided agents generate realistic and effective test events (RQ2.1), and whether their behaviors align with their defined interaction habits (RQ2.2). We analyze the quantity and quality of generated events, especially for input operations, to determine if different personas exhibit distinct and consistent action patterns that contribute to meaningful interaction diversity.

\textbf{RQ3} targets the \textbf{bug triggering capability} of persona-guided LLM agents. Beyond behavioral diversity, a primary goal of GUI testing is to identify app bugs. We evaluate how effectively the agents can uncover crash bugs (RQ3.1), typically triggered by invalid operations or edge-case inputs, as well as functional bugs (RQ3.2), such as misbehaving workflows, logic errors, or UI display issues. By analyzing the overlap situation of bugs triggered by each persona-guided LLM, we assess the practical benefits of incorporating human-like behavior diversity into LLM-based test automation.

We do not adopt code coverage as an evaluation metric because our testing tasks are scenario-specific and goal-driven, where coverage does not reliably indicate whether an agent meaningfully completes or exercises the target functionality \cite{yu2024practical}. In such settings, high coverage may result from unrelated interactions that do not contribute to the testing objective. Instead, we design our research questions to provide a comprehensive and task-aligned evaluation of \toolname, assessing the behavioral fidelity of personified agents, the validity of generated test events, the realism of input behaviors, and the ability to trigger functional and crash-inducing bugs. This multifaceted evaluation better reflects the effectiveness of emulating the diversity and intent of real-world crowdworkers in automated crowdsourced GUI testing.

To comprehensively evaluate our approach, RQ1–RQ3 are designed to capture complementary yet interrelated aspects of persona-guided testing. RQ1 examines whether different personas result in consistent and distinct exploration behaviors, validating the fidelity and interpretability of our structured persona modeling. It ensures that the injected personas meaningfully steer the agent's testing patterns. RQ2 then assesses the quality of actions produced along these paths, focusing on their contextual validity, executability, and functional soundness. While RQ1 focuses on how personas shape behavior, RQ2 verifies what those behaviors yield in terms of actionable tests. Building on both, RQ3 investigates whether persona-guided agents can effectively uncover real faults, demonstrating the practical impact of diverse exploration strategies. Together, these RQs establish that our approach not only produces distinct and interpretable behaviors (RQ1), but also maintains test reliability (RQ2) and achieves tangible testing value (RQ3).

\subsubsection{Data Preparation}

To ensure the generality and representativeness of our empirical evaluation, we collect a diverse set of 15 mobile apps as the subjects under test, balancing experimental generalization with practical resource constraints. The selected apps span a wide range of functional domains, including note-taking, shopping, travel, reading, \etc, to ensure representativeness. The included apps have varied GUI structures and interaction patterns. This broad domain coverage is intended to reduce domain-specific bias and to verify that \toolname generalizes across different usage flows.
These apps are selected from prior studies \cite{su2017guided, yu2024effective}, including 4 open-source and 11 commercial apps. Such apps are well-documented, publicly available, and widely studied apps with established experimental artifacts and realistic GUI complexity. This allows us to evaluate the effectiveness of \toolname for realistic and heterogeneous app conditions. To construct the evaluation tasks, we manually analyze the user interface, feature set, and documented usage patterns of each app through hands-on exploration and official app descriptions. This process involves identifying the primary user flows and high-frequency operations that represent the core functionality of the app \cite{yu2024practical}. Based on this analysis, we formulate a task per app that captures a realistic and representative user scenario, ensuring that the testing context aligns with actual usage behaviors (details on our online resources in \secref{sec:data}). We ensure that the evaluation reflects meaningful real-world interactions rather than abstract exploratory behaviors. This also allows us to assess the performance of each agent in contextually rich scenarios where human-like reasoning and persona alignment can play a significant role. Given the cost of multiple runs across different personas and baselines, scaling further is infeasible. 

\subsubsection{Baseline Construction}

We implement the baseline as a non-persona agent using the complete \toolname framework with the personification module disabled, while keeping GUI understanding, decision-making, and execution identical. This isolates the effect of persona injection and ensures a fair, controlled comparison. In cumulative experiments, we run the baseline nine times to match the number of personified agents, enabling equitable aggregation and reducing single-run variance. We also report one-to-one comparisons between each personified agent and the baseline to provide a fine-grained assessment.

In this work, we focus on task-oriented GUI testing, where agents are expected to complete specific functional scenarios rather than perform unguided exploration. Random testing strategies are not comparable, as they target whole-app exploration and lack semantic guidance aligned with predefined tasks \cite{mohammed2019empirical, yu2024practical, xiong2024general} (nevertheless, we run the random strategy to present the performance, and the data can be seen from \secref{sec:data}). Our baseline is a non-personified agent sharing the same decision-making process but without persona conditioning, allowing us to isolate the impact of structured persona injection. While \toolname adopts a similar LLM-based testing pipeline to prior approaches, a direct comparison is unnecessary, as our method is designed to be model-agnostic and pluggable. It can be integrated into existing frameworks, enhancing them with structured behavioral diversity rather than replacing or competing with them.

\subsubsection{Experimental Configuration}

To conduct a rigorous and reproducible evaluation of \toolname, we carefully configure the experimental environment. The architecture of \toolname relies on multiple LLM calls for different sub-tasks within the testing pipeline. Specifically, we use the GPT 4o model for GUI understanding and post-operation validation, due to its superior multimodal capabilities, which are essential for accurately interpreting screenshots and triggering GUI anomalies. The \texttt{temperature} is set to 0 for GPT 4o to ensure deterministic and consistent outputs. For the testing decision-making agent, we adopt the GPT o4-mini model, a lightweight and efficient LLM variant. This model is selected to simulate real-world usage conditions where computational cost is a concern, without significantly compromising reasoning ability. The \texttt{reasoning\_effort} is set to ``medium'' to balance response quality and latency. 

We use GPT-based models due to their strong performance in complex reasoning, contextual understanding, and multimodal capabilities, features critical for GUI testing tasks involving screen interpretation and intent generation. Our goal is to evaluate the effectiveness of persona-guided agent design, and using a state-of-the-art model ensures that observed outcomes reflect the method's potential rather than model limitations. Nonetheless, \toolname is model-agnostic and can be adapted to open-source LLMs as they continue to improve in accuracy and capability. In order to better illustrate the configurability, we replace the GPT models with DeepSeek models \cite{guo2025deepseek} to see the performance (see \secref{sec:data}). The results clearly support our conclusion that with different models, \toolname can show obvious intra-cluster cohesion and inter-cluster separation.

To account for the stochastic nature of LLM-based decision-making and to capture potential behavioral patterns, each task is executed five times per agent configuration, each time with a 20-minute execution. In order to determine the execution time, we recruit five graduate students who command the necessary and basic knowledge for crowdsourced GUI testing (as representatives for common crowdworkers \cite{salman2015students}), and the average time of each of them to finish each task is around 10 minutes. We double the time for LLM agents and set the limit to 20 minutes. This repetition allows us to observe intra-persona consistency and enables the computation of statistically robust metrics for evaluation. Running multiple sessions per task also provides insight into whether the agents demonstrate stable behavioral traits aligned with their assigned personas across executions.

\subsection{RQ1: Exploration Trend}

To assess whether \toolname exhibits exploration behaviors that are consistent within the same persona and diverse across different personas, we design a comprehensive evaluation consisting of three components: intra-cluster cohesion (RQ1.1), inter-cluster separation (RQ1.2), and persona consistency (RQ1.3). This evaluation enables us to determine whether the behavioral differences encoded in personas are faithfully reflected in the test paths produced by the agents.

For RQ1.1, intra-cluster cohesion quantifies the degree to which repeated test executions of the same persona result in similar exploration trajectories. For each persona and each task, we conduct five independent test runs. The similarity between each pair of paths is computed, and their average is reported as the cohesion score. The cohesion is calculated as the mean pairwise similarity across all $n \times (n-1) / 2$ combinations, where $n = 5$ denotes the number of runs: $\text{Cohesion}_{\text{intra}} = \frac{\sum_{i=1, j=1, i<j}^{n} \text{sim}_{i,j}}{n}$. For RQ1.2, inter-cluster separation measures the distinctiveness between different personas by computing the average similarity between paths generated by different persona-guided agents. Specifically, for each pair of personas $M$ and $N$, we compare every test trace from $M$ with every trace from $N$, and calculate the average of all $5 \times 5 = 25$ pairwise similarities: $\text{Separation}_{\text{inter}}(M, N) = \frac{\sum_{i=1}^{5} \sum_{j=1}^{5} \text{sim}_{i,j}}{25}$. A high intra-cluster cohesion combined with low inter-cluster similarity indicates that each persona induces stable yet distinct exploration behavior.

\begin{wrapfigure}{l}{0.6\textwidth}
\centering
\vspace{-10pt}
\includegraphics[width=\linewidth]{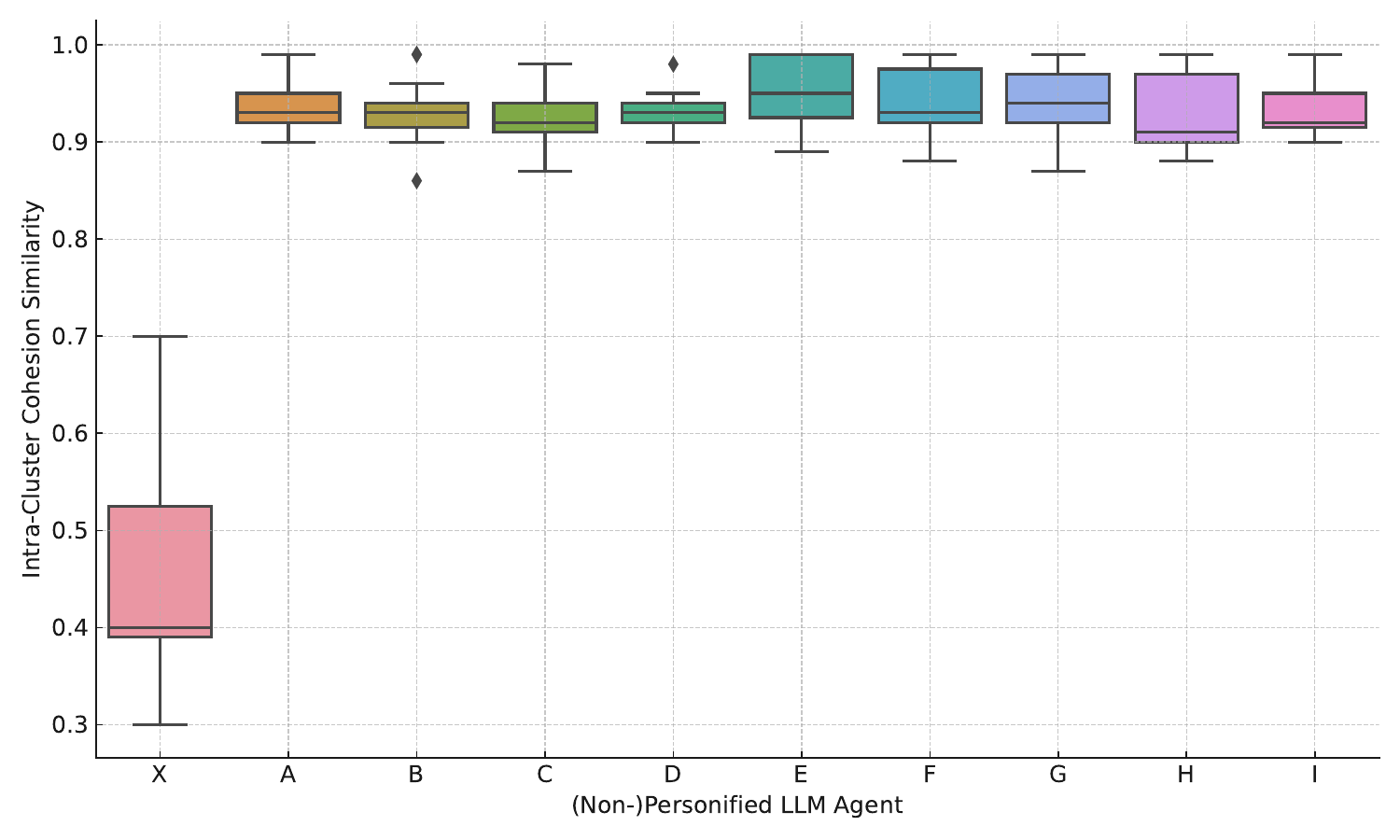}
\vspace{-15pt}
\caption{RQ1.1: Intra-Cluster Cohesion}
\vspace{-10pt}
\label{fig:rq11}
\end{wrapfigure}

To compute the similarity between two test paths, we first encode each path as a vector and then measure the cosine similarity between them. The path vectorization process is a multi-step procedure designed to capture semantic-level behavioral patterns in a principled and interpretable manner. First, we validate the exploration traces to construct cleaned and interpretable test traces. To transform raw GUI interactions into concise natural language phrases, we apply a lightweight, rule-based purification process \cite{yu2021prioritize}. Each interaction is first mapped to its action type (\eg click, input) based on execution logs, and the corresponding widget label is extracted from GUI metadata and OCR results. These elements are then combined using simple verb-object templates (\eg ``click save button'', ``input alarm time'') \cite{yu2021prioritize, wang2019images}, with fallback descriptions for unlabeled widgets. The resulting phrases are normalized for consistency and subsequently encoded using SBERT, providing a semantically meaningful and scalable representation of exploration traces. Each action phrase is individually encoded using Sentence-BERT (SBERT) \cite{reimers2019sentence}, which converts the sentence into a 384-dimensional vector representation. The entire sequence of encoded actions is then passed through a bidirectional LSTM (BiLSTM) model \cite{graves2013hybrid}, which aggregates the sequence into a fixed-size 256-dimensional vector representing the full test path. We choose SBERT because it provides semantically meaningful embeddings for short natural language phrases, making it well-suited for encoding individual GUI actions. To capture the sequential and contextual structure of entire test paths, we use a BiLSTM, which effectively models both forward and backward dependencies in action sequences. This combination ensures that both semantic intent and behavioral flow are preserved in the final path representation. This model configuration, including SBERT and BiLSTM, is selected based on GPT-assisted parameter tuning and prior studies in semantic sequence encoding. Cosine similarity is computed over path-level embeddings generated by a BiLSTM, which aggregates SBERT-encoded action phrases in sequence. Each action phrase is produced through a rule-based purification process that abstracts low-level GUI events into concise, semantically meaningful descriptions. The BiLSTM captures both the semantic content and the sequential dependencies of these actions, ensuring that variations in execution order and context are preserved in the final embedding. Applying cosine similarity to these embeddings allows us to quantify differences in exploration behavior, capturing not just which actions were taken but also how they were structured. This design aligns with our goal of evaluating persona-induced behavioral patterns, supporting the measurement of intra-persona cohesion and inter-persona separation in a principled and order-sensitive manner.

To prepare for the user study evaluating persona consistency (RQ1.3), we recruited 20 participants with relevant experience in software GUI testing, including graduate students specializing in software engineering and professional QA engineers or testers with over five years of industry experience. For the study materials, we generated a set of complete video recordings capturing GUI exploration traces produced by different persona-guided LLM agents across various tasks. All videos were carefully anonymized to remove any labels or identifiers that might reveal the underlying persona. After viewing each video, participants were provided with a list of predefined persona profiles, each described along three dimensions: testing mindset, exploration strategy, and interaction habit. Participants were asked to evaluate the degree to which the observed agent behavior aligned with each persona profile using a 10-point Likert scale \cite{joshi2015likert}, where 1 indicated no match and 10 indicated a perfect match. This design allowed us to quantitatively assess the interpretability and semantic fidelity of the persona-driven behaviors from a human perspective.

For the intra-cluster cohesion, we quantify the similarity between multiple test sessions executed by the same persona-guided agent on the same task. As shown in \figref{fig:rq11}, most personified agents demonstrate high intra-persona consistency, with scores often approaching or exceeding 0.9. This indicates that the decision-making process of each agent is stably guided by its assigned persona, resulting in reproducible behavior patterns across repeated runs. In contrast, the non-personified baseline agent (\ps{X}) exhibits noticeably lower cohesion, ranging from 0.31 to 0.7, reflecting less stable and more erratic exploration behavior in the absence of explicit behavioral modeling. These findings confirm that persona injection not only informs the high-level intent of an agent but also anchors its interaction behavior in a consistent and repeatable manner.

\begin{wrapfigure}{l}{0.7\textwidth}
\centering
\vspace{-10pt}
\includegraphics[width=\linewidth]{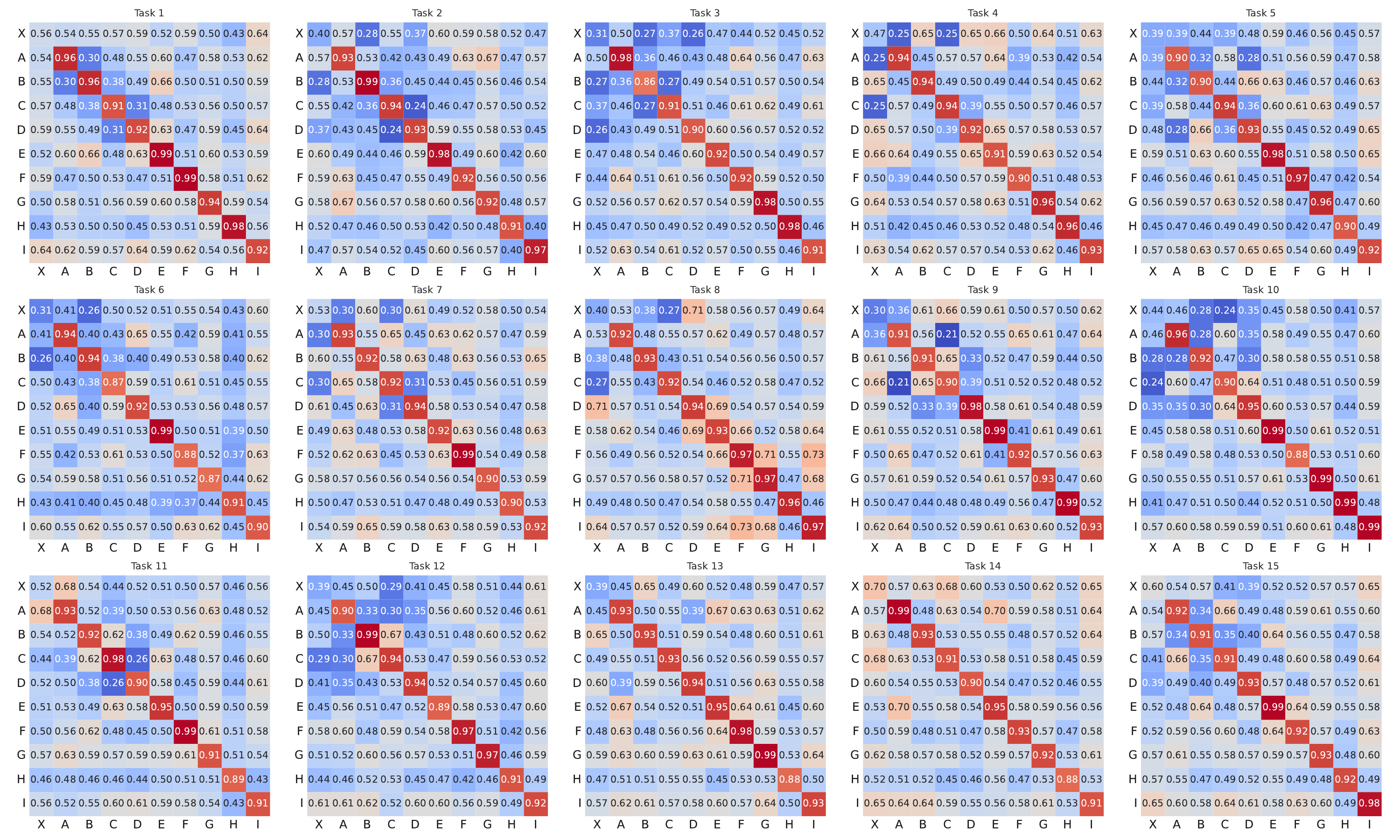}
\vspace{-20pt}
\caption{RQ1.2: Inter-Cluster Separation}
\vspace{-10pt}
\label{fig:rq12}
\end{wrapfigure}


To assess behavioral distinctiveness across personas, we measure inter-cluster separation among agents executing the same tasks. As shown in \figref{fig:rq12}, pairwise similarity matrices reveal that inter-persona similarities are consistently far lower than intra-persona ones, confirming that each persona drives distinct exploration behavior. For instance, although \ps{B} and \ps{C} share a sequential mindset, their differing strategies and input habits result in notable behavioral divergence, with similarity scores ranging from 0.27 to 0.67. Likewise, divergent personas like \ps{E} and \ps{G} show clear separation from structured ones like \ps{A} and \ps{B}. These results highlight the effectiveness of our persona configuration space in producing diverse, behaviorally meaningful testing patterns.

For RQ1.3, the results indicate a high degree of perceived consistency between the persona-guided agent behaviors and their intended persona profiles. The average ratings across all participants are 8.35 for Testing Mindset, 8.70 for Exploration Strategy, and 8.80 for Interaction Habit (all scores are over or equal to 7). These high scores suggest that participants were able to consistently recognize and interpret the distinct behavioral characteristics encoded in the personas.

Among the three dimensions, Interaction Habit achieves the highest average rating (8.80) with relatively low standard deviation (0.77), indicating that participants found the input style and behavioral tendencies (\eg inputting valid and long or invalid values) particularly distinguishable and reliably expressed. Testing Mindset follows closely with an average score of 8.35 and the lowest standard deviation (0.67), showing stable agreement across participants in identifying sequential or divergent thinking patterns in agent behavior. Exploration Strategy, while still receiving strong ratings (average of 8.70), exhibits the highest variability with a standard deviation of 1.08. This suggests that although participants generally perceived exploration intent (\eg click-focused or core-function-focused) as aligned with the persona descriptions, such strategies may be somewhat more subjective or less consistently expressed compared to the other dimensions.

These results validate that the behaviors produced by persona-guided LLM agents are not only diverse and consistent but also interpretable and semantically meaningful from a human perspective. The strong alignment between agent behavior and persona descriptions confirms the effectiveness of \toolname in simulating realistic and distinguishable crowdworker-like testing styles.

\subsection{RQ2: Test Generation Effectiveness}

\begin{figure}[!h]
\centering
\begin{subfigure}[b]{0.48\linewidth}
    \centering
    \includegraphics[width=0.9\linewidth]{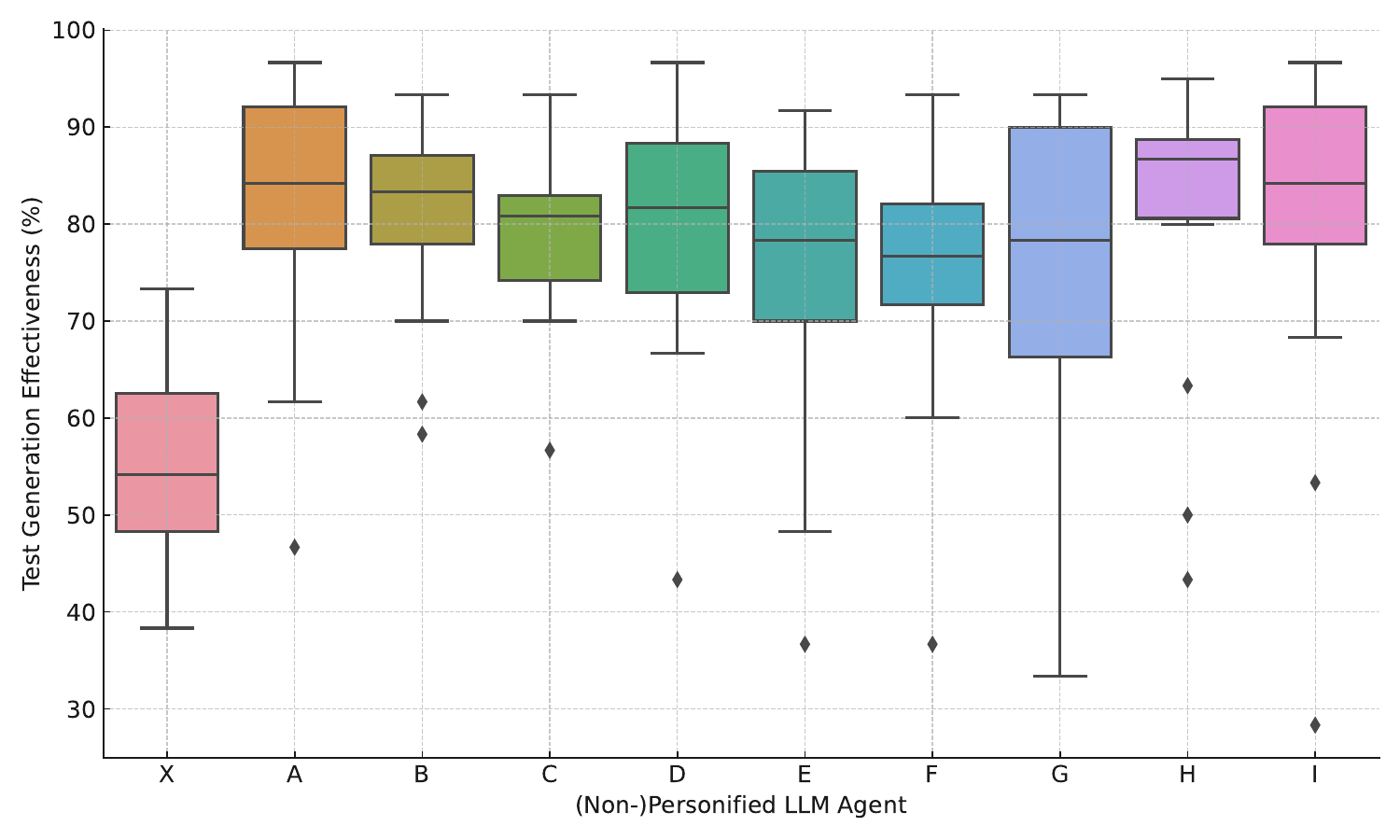}
    \vspace{-5pt}
    \caption{RQ2.1: General Test Generation Effectiveness}
    \label{fig:rq21}
\end{subfigure}
\hfill
\begin{subfigure}[b]{0.48\linewidth}
    \centering
    \includegraphics[width=0.9\linewidth]{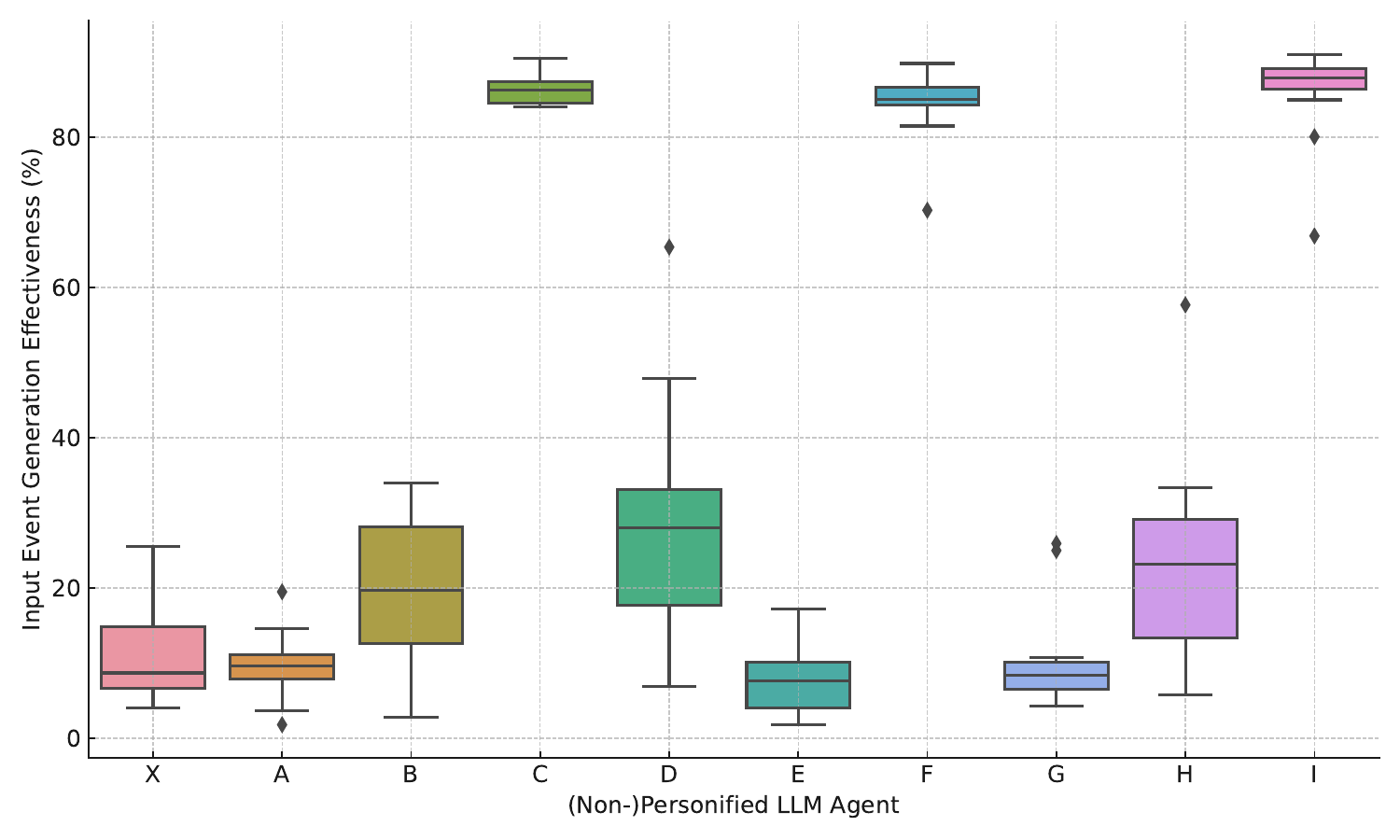}
    \vspace{-5pt}
    \caption{RQ2.2: Input Test Generation Effectiveness}
    \label{fig:rq22}
\end{subfigure}
\vspace{-5pt}
\caption{RQ2: Test Generation Effectiveness}
\vspace{-10pt}
\label{fig:rq2}
\end{figure}

RQ2 investigates the effectiveness of persona-guided LLM agents in generating test events during automated GUI exploration. We evaluate this effectiveness from two perspectives: general test event generation, which includes all interaction types such as clicks and navigations, and input event generation, which focuses specifically on text-based input actions like entering values into forms or fields. In this study, we define test generation effectiveness as the degree to which a generated test event contributes meaningfully to the exploration of the app under test. Specifically, we consider a test event to be effective if it leads to a successful and semantically valid interaction, such as triggering a new GUI state, invoking a meaningful response, or progressing a task scenario. We calculate the proportion of effective events to all events and show the results in \figref{fig:rq2}. 

Most persona-guided agents perform comparatively better than the non-personified baseline agent (\ps{X}) in terms of overall test event generation effectiveness, outperforming by 33\% -- 47\% on average. Several agents consistently outperform the baseline. This indicates that the integration of persona-guided behavioral intent does not compromise the LLM's ability to interact with the GUI. On the contrary, it appears to guide the agent toward producing more purposeful and executable test actions.

A more distinct pattern is observed in input event generation effectiveness, presented in \figref{fig:rq22}. The results clearly demonstrate that this capability is strongly influenced by the exploration strategy specified in the persona. Agents with input-oriented strategies (\ie \ps{C}, \ps{F}, \ps{I}) achieve consistently high effectiveness scores across multiple tasks, outperforming the baseline \ps{X} by 683.32\% -- 697.50\% on average. Their behaviors are aligned with a focused intent to interact with input fields, resulting in a higher frequency of valid and semantically appropriate text inputs. In contrast, agents with click-oriented exploration strategies (\ie \ps{A}, \ps{E}, \ps{G}) consistently exhibit low input event effectiveness, outperforming the baseline \ps{X} by -28.03\% -- -6.60\% on average. These agents prioritize interaction with clickable UI elements and generally ignore input components, as specified in their persona definitions. Agents with core-function-focused strategies (\ie \ps{B}, \ps{D}, \ps{H}) show more variable performance, outperforming the baseline \ps{X} by 80.31\% -- 156.59\% on average. While they occasionally produce valid input events during task-oriented workflows, their effectiveness in this area remains inconsistent and generally moderate.

These results confirm that the persona design can directly shape the LLM agent's attention and interaction patterns. The strong alignment between input-oriented strategies and input event generation effectiveness demonstrates the importance of fine-grained behavioral modeling in achieving targeted test coverage. Overall, the findings support the conclusion that persona-guided agents not only enhance diversity and realism in automated GUI testing, but also contribute to the generation of more meaningful and functional test events.

\subsection{RQ3: Bug Triggering Capability}

RQ3 evaluates the practical effectiveness of persona-guided LLM agents in triggering bugs during automated GUI testing. Specifically, we focus on two categories of bugs: crash bugs and functional bugs. The goal is to determine whether the diversity introduced through personification contributes meaningfully to bug triggering beyond what a single non-personified agent can achieve.

\begin{figure}[!h]
\centering
\vspace{-10pt}
\includegraphics[width=0.75\linewidth]{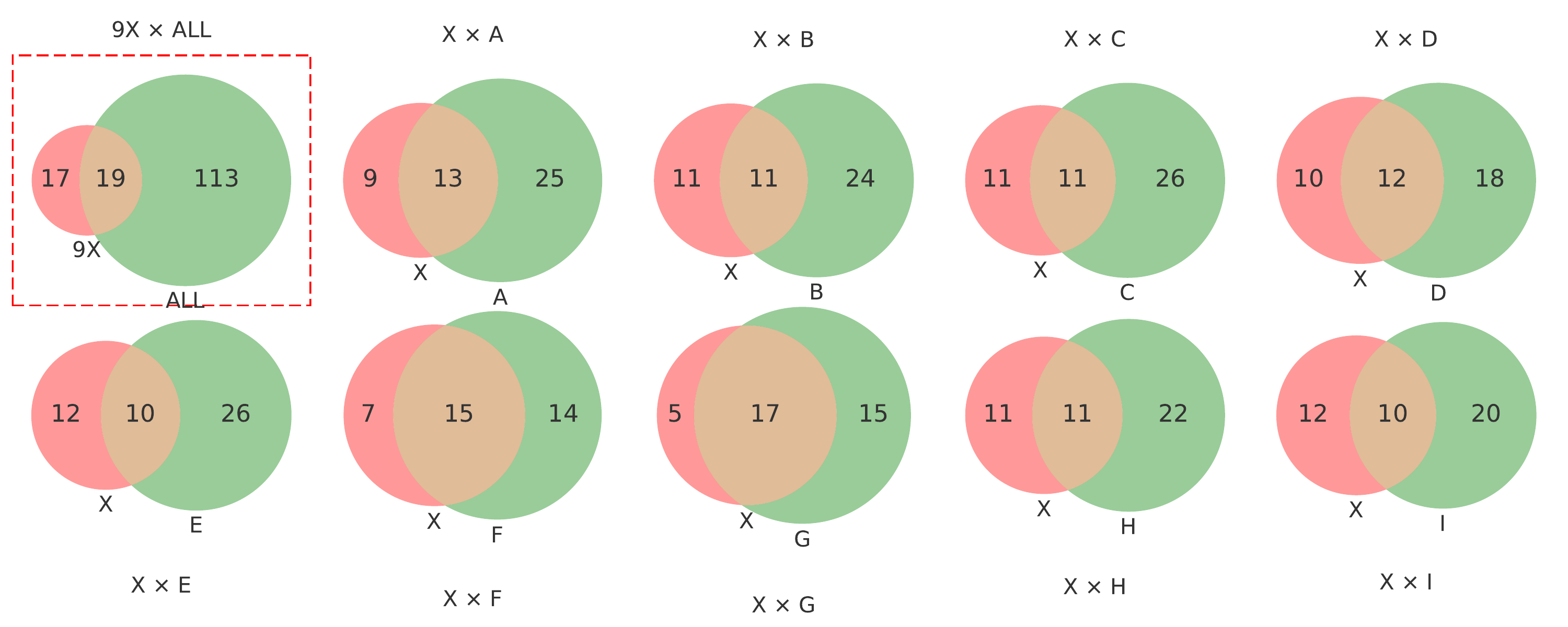}
\vspace{-10pt}
\caption{RQ3.1: Crash Bug Triggering Capability}
\vspace{-10pt}
\label{fig:rq31}
\end{figure}

Crash bug triggering results are summarized in \figref{fig:rq31}. Each subfigure presents a comparison between the set of crash bugs triggered by the non-personified agent, \ps{X}, and each persona-guided agent (\ps{A} through \ps{I}). The overlapping regions illustrate the shared bug findings, while non-overlapping regions highlight unique discoveries. Across all comparisons, persona-guided agents demonstrate an ability to trigger both common and distinct crash bugs relative to the baseline. Six of the agents trigger over 20 distinct bugs that are not covered by the baseline \ps{X}, among which \ps{E} has the best performance, triggering 26 distinct bugs. Intuitively, persona-guided agents trigger 29 -- 38 crash bugs in total on different tasks while \ps{X} can only trigger 22 crash bugs. This suggests that the exploratory diversity introduced through persona configurations enables agents to reach UI states or usage paths that are less accessible to a generic strategy.


We observe that a few bugs are uniquely triggered by the non-personified baseline agent. These cases are largely due to the inherent randomness in the baseline's decision-making, which occasionally leads to alternative, unstructured paths that are not aligned with any defined persona. However, such behavior is non-repeatable and shows no consistent pattern. In contrast, persona-guided agents exhibit stable, interpretable behaviors. While our current persona set is not exhaustive, these findings suggest that expanding the diversity of personas can further improve bug coverage. Overall, the results highlight that personified agents can uncover unique bugs aligned with their behavioral profiles, validating the value of structured diversity in automated GUI testing.

To further assess the practicality, we compare the cumulative results of nine persona-guided agents (\ps{A} through \ps{I}) against a nine-run result of the \ps{X}, as shown in subfigure ``9X $\times$ ALL''. In this setting, we simulate the scenario of replacing nine human crowdworkers with nine diverse persona-driven agents. The results indicate that the persona-guided group collectively discovers a significantly broader set of crash bugs. While there is some overlap, the union of unique bugs found by the persona agents substantially exceeds that of nine repetitions of the baseline. This confirms that behavioral diversity, rather than repetition, is the key factor in maximizing crash bug exposure.

Functional bug triggering results are presented in \tabref{tab:rq32}. The table lists functional bugs triggered by each agent across a set of predefined tasks. Each entry corresponds to a specific bug triggered by an agent in a particular task. Bug IDs in \tabref{tab:rq32} are assigned based on matched failure symptoms and triggering conditions. If multiple agents encounter the same issue under similar interaction paths and GUI states, their findings are grouped under a single ID. This process ensures that cross-agent overlaps reflect actual shared bug discoveries rather than superficial similarity.

All reported functional bugs are confirmed through manual inspection. Three of the authors independently review the exploration traces, including UI context, executed actions, and resulting outcomes. A functional bug is labeled only when all reviewers reach consensus, ensuring consistent and objective validation. The multimodal LLM is used solely for analyzing GUI-level visual issues (\eg layout errors), not for determining functional correctness.

Among all tasks, \ps{H} achieves the highest triggering count (6 bugs), followed by \ps{B} and \ps{G} (4 bugs each). Interestingly, the non-personified agent only triggers 3 bugs in total, which are all triggered by some of the persona-guided agents. These results reveal a similar trend to crash bug triggering: persona-guided agents can not only discover the same set of functional bugs as the baseline but also contribute to triggering additional, previously unseen issues. 

\begin{table}[!htbp]
\vspace{-10pt}
\caption{RQ3.2: Functional Bug Triggering Capability}
\vspace{-10pt}
\label{tab:rq32}
\resizebox{0.95\columnwidth}{!}{\begin{tabular}{ccccccccccc}

\toprule
Task ID & \ps{X} & \ps{A} & \ps{B} & \ps{C} & \ps{D} & \ps{E} & \ps{F} & \ps{G} & \ps{H} & \ps{I} \\ \midrule

4  & fb11 & fb11 & fb11 & fb11 & fb11 & fb11 & & & fb11 & \\
5  & & & fb10 & fb10 & fb10 & & & & & fb10 \\
6  & & & & & & fb1 & fb1 & fb1 & fb1 & fb1 \\
9  & fb6 & fb7 & & & & & fb7 & fb9 & fb6 fb7 fb8 & fb8 \\
10  & & fb2 & fb3 & & & & & fb2 & & \\
15  & fb4 & & fb4 & fb4 & & fb4 & fb5 & fb4 & fb4 & \\ \midrule
SUM & 3 & 3 & 4 & 3 & 2 & 3 & 3 & 4 & 6 & 3 \\ \bottomrule
\end{tabular}}
\end{table}

To illustrate the influence of persona configurations on functional bug triggering, we highlight some representative cases where the behavioral traits of persona-guided agents directly contribute to the discovery of functional bugs.

\textbf{fb2 (Timing-Sensitive Filtering Error)} is triggered only when the alarm filter is applied at the very start of the session. This bug is discovered by agents with \ps{A} and \ps{G}, both of which adopt the sequential testing mindset and click-oriented strategy that prioritize early interaction with top-level UI elements. Their methodical behavior aligns precisely with the temporal condition required to expose this transient issue.

\textbf{fb3 (Alarm Editing Lock-In) \textit{(the bug presented in \secref{sec:mv})}} leads to an inactive alarm to become uneditable after a type change. This is triggered exclusively by the agent using \ps{B}, which emphasizes coherent, goal-driven exploration. The persona's complete workflow execution, including editing and saving, is essential to reach the bug state.

\textbf{fb5 (Peripheral Page Rendering Bug)} occurs on a rarely visited ``Support Development'' page, where lower content fails to render. \ps{F}, designed with divergent and input-seeking traits, navigates through non-obvious paths while searching for input opportunities, uncovering a bug that task-focused strategies would likely miss.

\textbf{fb7 (Text-Length UI Overlap)} involves a visual flaw where the ``Send'' button is obscured by long text input. Agents with \ps{A}, \ps{F}, \ps{H}, all configured with long-valid input interaction habit, consistently surface this bug. Their interaction habit simulates realistic user stress conditions, revealing UI vulnerabilities missed by short or default inputs.

These examples demonstrate that persona traits such as interaction timing, input style, and exploration breadth have a tangible impact on the types of bugs discovered. By encoding realistic user behaviors, \toolname enables automated agents to uncover defects that align with real-world usage patterns. In summary, the findings from RQ3 confirm that persona-guided LLM agents enhance the bug triggering capability of automated GUI testing. By embedding diverse behavioral intents into the agents, \toolname simulates a virtual crowd of testers whose collective exploration patterns lead to higher bug triggering than generic or repetitive automation. This validates the practical benefit of incorporating behavioral diversity in automated testing, particularly in settings where maximizing bug discovery is critical.

\subsection{Threats to Validity}

One threat to internal validity stems from the manual labeling of exploration traces used in the similarity analysis and user study. To reduce bias, we applied cross-validation and used standardized linguistic templates for action descriptions. The user study also employed randomized trace presentation and anonymized persona identities to minimize experimenter and participant bias. External validity may be limited by our selection of test apps, tasks, and the defined personas. While these cover diverse behavioral patterns and real-world tasks, they may not capture the full spectrum of user behaviors or application domains. To mitigate this, we varied task types and UI structures and grounded persona design in real-world crowdworker data. Measurement validity also presents a risk, as metrics like path similarity may not fully capture complex behavioral nuances. To address this, we complement quantitative analysis with qualitative assessments, including a user study to evaluate semantic alignment, ensuring a more robust validation of persona-driven testing fidelity. We have tried our best to design a rigorous evaluation protocol that balances reproducibility, interpretability, and realism. These steps collectively strengthen the validity of our findings and support the broader applicability of persona-guided LLM agents in automated GUI testing.

\subsection{Discussion} 

\subsubsection{Novelty Highlight} 

A key contribution of \toolname is how we simulate crowdsourced testing through a structured, automated approach. Rather than replicating the same agent multiple times or relying on implicit behavioral randomness, we inject diverse personas into a shared LLM-based testing framework. This allows personified agents to represent different crowdworker archetypes, and enables them to work collectively to emulate the behavioral diversity of the real-world crowdsourced testing paradigm. The resulting ensemble of persona-guided agents produces complementary testing behaviors that, together, achieve broader exploration and more effective bug triggering than any single agent alone. To sum up, \toolname can be positioned as an automated realization of crowdsourced GUI testing.

\subsubsection{Influence of App Characteristics} 

The effectiveness can be influenced by factors such as GUI and functional complexity, UI frameworks, and application domain. While a detailed factor-wise analysis is beyond the scope of this study, we mitigate potential bias by selecting apps from diverse domains with varied GUI structures and interaction patterns, implicitly covering different levels of complexity. We observe consistent trends across these heterogeneous apps, suggesting that persona-guided agents are robust to such variations. A more fine-grained analysis of how specific app characteristics affect persona-guided testing remains an important direction for future work.

\subsubsection{Addressing LLM Inherent Limitations}

\toolname mitigates inherent LLM limitations such as uncertainty, hallucination, and reproducibility through explicit design choices. A two-stage intent verification mechanism validates each proposed action against the current GUI context using an independent LLM-as-a-judge, reducing misaligned decisions and preventing error propagation. In addition, deterministic LLM settings (\eg \texttt{temperature} = 0) are used to ensure stable and reproducible behavior across runs, enabling fair evaluation of persona effects.

\subsubsection{Economic Cost Analysis}

\toolname adopts a modular model configuration to balance accuracy and efficiency. The lightweight o4-mini model is used for decision-making, while the more capable multimodal GPT-4o model is reserved for GUI understanding and post-execution validation, where higher precision is needed. Based on our measurements, GUI understanding and validation consume, on average, 112,386 input tokens (\textasciitilde \$0.28) and 12,305 output tokens (\textasciitilde \$0.12) per task per agent. The decision-making process consumes 70,569 input tokens (\textasciitilde \$0.08) and 45,720 output tokens (\textasciitilde \$0.20). In total, each task execution costs \textasciitilde \$0.68 per agent, which is more economical than recruiting a real crowdworker to perform the same testing task. The breakout tables for the results are on our online supplementary materials (\secref{sec:data}) due to page limit. This cost-effectiveness enables scalable deployment of PersonaTester for automated crowdsourced testing.

\section{Related Work}


\subsection{LLM-based GUI Testing: Automated and Crowdsourced}

Recent studies have explored leveraging LLMs to augment automated GUI testing. Liu \etal \cite{liu2024make} propose GPTDroid, which passes the app GUI state to an LLM and iteratively ``chats'' with the app to generate test actions. Wang \etal \cite{wang2025llmdroid} introduce LLMDroid, which combines traditional automated exploration with occasional LLM guidance. Overall, these studies demonstrate that LLMs can drive human-like exploratory testing, strategically navigating app GUI and increasing coverage, if used judiciously to balance insight versus cost. Beyond raw coverage, researchers are using LLMs to pursue more semantic testing goals aligned with app functionality and user scenarios. Yu \etal comprehensively analyze the capabilities of LLMs in generating and migrating GUI test scripts \cite{yu2023llm}. Yoon \etal develop DroidAgent \cite{yoon2024intent}, an autonomous LLM-driven tester that sets high-level task goals and then executes the GUI steps to fulfill them. Similarly, Yu \etal successively present ScenTest \cite{yu2024practical} and ScenGen \cite{yu2025llm}, which are frameworks that mirror the manual testing process by assigning different sub-tasks to LLM agents or knowledge graph enhanced modules. 

LLMs have also been applied to migrate GUI test scripts across different apps or platforms. Traditional test migration techniques relied on static widget mappings between apps with similar functionality, but this often breaks down when the GUI diverges \cite{zhang2024synthesis}. Zhang \etal \cite{zhang2024llm} propose an abstraction-concretization paradigm using LLMs. Another approach by Cao \etal \cite{cao2025intention} uses LLM ``intention understanding'' to migrate tests. These studies illustrate how LLMs serve as software testing translators, carrying out test reuse by understanding the purpose behind GUI actions.

Another promising application is using LLMs to interpret bug reports and reproduce the described issues on the GUI. AdbGPT \cite{feng2024prompting} is an early system that employs GPT 3.5 to extract structured S2R steps from a free-form bug report, then iteratively uses ChatGPT to select matching GUI widgets and execute each step. Wang \etal \cite{wang2024feedback} introduce ReBL, which forgoes explicit step extraction and instead feeds the entire bug description to the LLM, coupled with a feedback loop during execution. Another effort is ReActDroid by Huang \etal \cite{huang2025one}, which combines ``Reasoning + Acting'' prompts to infer missing context and steps from just the crash summary. These advances show that LLMs can serve as powerful natural language intermediaries in testing: they read and comprehend bug descriptions or logs, then drive the app GUI to reenact failures for developers.

LLMs can also be utilized to enhance the crowdsourced testing, including task allocation \cite{xie2017cocoon, cui2017should, wang2019characterizing}, report clustering \cite{hao2019ctras, chen2021effective, liu2020clustering, yu2024semi}, report duplication detection \cite{wang2019images, kim2022predicting, wu2023intelligent}, prioritization \cite{tong2021crowdsourced, yang2021crowdsourced, fang2024enhanced}, quality detection \cite{yu2023mobile}, \etc LLMPrior \cite{ling2025redefining}, a representative LLM-based framework, prioritizes crowdsourced bug reports by leveraging semantic understanding. However, no existing work has aimed to enhance crowdsourced testing by automating the behavior of crowdworkers; current methods still rely heavily on human participation and its inherent unpredictability. This paper introduces \toolname, a personified-LLM-based framework that simulates diverse tester behaviors through designed personas, thereby improving the efficiency, reliability, and controllability of crowdsourced testing without sacrificing behavioral diversity.

\subsection{LLM Personification}

Another active research area is LLM personification, injecting personas into LLMs to steer behavior and enhance task performance. Originally explored in conversational AI (\eg PersonaChat), personification has gained traction as a powerful prompt-engineering strategy across domains \cite{tseng2024two}. Role-based prompts (\eg ``act as a software tester'') influence the LLM's reasoning style and focus.

Recent studies suggest that well-aligned personas can improve task outcomes. For instance, Hu \etal \cite{hu2024quantifying} find that LLM responses vary notably in tone and content across 162 tested personas, with benefits dependent on persona-task alignment and prompt formulation. In software engineering, prompting LLMs as testers or developers leads to more relevant and focused responses. Multi-agent frameworks have also embraced persona specialization, \eg Luo \etal \cite{luo2024personamath} and Qian \etal \cite{qian2023communicative} demonstrate that role-assigned LLM agents (planner, coder, tester) achieve better collaboration and structured reasoning. These findings collectively highlight persona injection as a practical means for achieving more effective and targeted LLM behavior.

Beyond NLP tasks, LLM personification gains traction in HCI for simulating user personas. Sun \etal \cite{sun2025persona} introduce Persona-L, which uses ability-based persona prompts to model users with complex needs, such as individuals with disabilities. In social and educational simulations, Park \etal \cite{park2023generative} present ``generative agents'', LLMs with personalities that interact autonomously in virtual environments, exhibiting human-like behaviors. These studies highlight the versatility of persona-based techniques, showing the ability to steer LLM outputs for human-centered reasoning. LLM personification is a versatile method for guiding an LLM's knowledge, style, and reasoning, enhancing its effectiveness for specific roles or user needs.

\section{Conclusion}

This paper presents \toolname, a novel framework that integrate persona into LLM-driven agents to automates crowdsourced testing. By modeling testers across three dimensions (testing mindset, exploration strategy, interaction habit), \toolname enables realistic and diverse test behaviors. Experimental results show strong intra-persona consistency and clear inter-persona differentiation, effectively replicating real-world crowdworker patterns. Compared to baseline agents, \toolname improves test generation effectiveness and bug triggering rates, showing its potential to bridge the gap between manual and automated testing paradigms.

\section{Data Availability}
\label{sec:data}

More details and the replication package are on \textbf{\url{https://sites.google.com/view/personatester}}.

\begin{acks}
Shengcheng Yu is supported partially by the Institute for Advanced Study (IAS) of TUM, Zhenyu Chen is supported partially by the National Key Research and Development Program of China (2024YFF0908001).
\end{acks}

\bibliographystyle{ACM-Reference-Format}
\bibliography{main}

\end{document}